\documentclass[aps,twocolumn,prb,superscriptaddress,showpacs,amsmath,a4paper]{revtex4}
\usepackage{graphics,graphicx,color}
\begin{document}
\title{Dzyaloshinskii-Moriya interaction and spin re-orientation transition in the frustrated kagome lattice antiferromagnet}
\author{K.~Matan}
	\email{sckittiwit@mahidol.ac.th}
 	\affiliation{Department of Physics, Massachusetts Institute of Technology, Cambridge, MA 02139, USA} 
 	\affiliation{Neutron Science Laboratory, Institute for Solid State Physics, University of Tokyo, 106-1 Shirakata, Tokai, Ibaraki 319-1106, Japan}
	\affiliation{Department of Physics, Mahidol University, 272 Rama VI Rd., Ratchathewi, Bangkok 10400, Thailand.}	
\author{B.~M.~Bartlett}
	\affiliation{Department of Chemistry, Massachusetts Institute of Technology, Cambridge, MA 02139, USA} 
	\affiliation{Department of Chemistry, University of Michigan, Ann Arbor, MI 48109, USA}
\author{J.~S.~Helton}
	\affiliation{Department of Physics, Massachusetts Institute of Technology, Cambridge, MA 02139, USA} 
	\affiliation{NIST Center for Neutron Research, National Institute of Standards and Technology, Gaithersburg, MD 20899, USA}
\author{V.~Sikolenko}
	\affiliation{Paul Scherrer Institute, 5232 Villigen PSI, Switzerland} 
\author{S.~Mat\/'a\v{s}}
	\affiliation{Helmholtz-Zentrum Berlin f\"ur Materialien und Energie GmbH, Hahn-Meitner-Platz 1, D-141 09 Berlin, Germany}
\author{K.~Proke\v{s}}
	\affiliation{Helmholtz-Zentrum Berlin f\"ur Materialien und Energie GmbH, Hahn-Meitner-Platz 1, D-141 09 Berlin, Germany}
\author{Y.~Chen}
	\affiliation{NIST Center for Neutron Research, National Institute of Standards and Technology, Gaithersburg, MD 20899, USA}
\author{J.~W.~Lynn}
	\affiliation{NIST Center for Neutron Research, National Institute of Standards and Technology, Gaithersburg, MD 20899, USA}
\author{D.~Grohol}
	\affiliation{The Dow Chemical Company, Core R\&D, Midland, MI 48674, USA}
\author{T.~J.~Sato}
	\affiliation{Neutron Science Laboratory, Institute for Solid State Physics, University of Tokyo, 106-1 Shirakata, Tokai, Ibaraki 319-1106, Japan}
\author{M.~Tokunaga}
	\affiliation{Institute for Solid State Physics, University of Tokyo, 5-1-5 Kashiwanoha, Kashiwa, Chiba 277-8581, Japan}
\author{D.~G.~Nocera}
	\affiliation{Department of Chemistry, Massachusetts Institute of Technology, Cambridge, MA 02139, USA}
\author{Y.~S.~Lee}
	\email{younglee@mit.edu} 
	\affiliation{Department of Physics, Massachusetts Institute of Technology, Cambridge, MA 02139, USA}
\date{\today}

\begin{abstract}

Magnetization, specific heat, and neutron scattering measurements were performed to study a magnetic transition in jarosite, a spin-$\frac{5}{2}$ kagome lattice antiferromagnet. When a magnetic field is applied perpendicular to the kagome plane, magnetizations in the ordered state show a sudden increase at a critical field $H_c$, indicative of the transition from antiferromagnetic to ferromagnetic states. This sudden increase arises as the spins on alternate kagome planes rotate 180$^\circ$ to ferromagnetically align the canted moments along the field direction. The canted moment on a single kagome plane is a result of the Dzyaloshinskii-Moriya interaction.  For $H<H_c$, the weak ferromagnetic interlayer coupling forces the spins to align in such an arrangement that the canted components on any two adjacent layers are equal and opposite, yielding a zero net magnetic moment.  For $H>H_c$, the Zeeman energy overcomes the interlayer coupling causing the spins on the alternate layers to rotate, aligning the canted moments along the field direction. Neutron scattering measurements provide the first direct evidence of this 180$^\circ$ spin rotation at the transition.

\end{abstract}

\pacs{75.25.-j, 75.30.Kz, 75.50.Ee, 75.40.Cx} 
\maketitle
\section{Introduction}

\begin{figure}[t]
\centering
\includegraphics[width=2.8in]{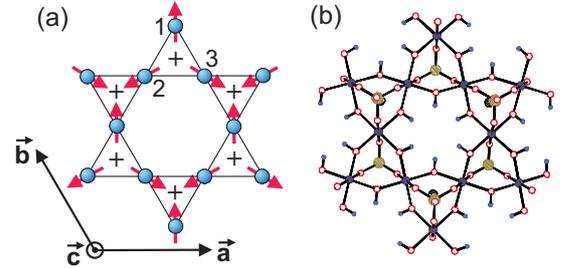}
\caption{(Color online) (a)~The kagome lattice is drawn as corner-sharing triangles with the $q=0$ ground state spin structure. Plus signs inside triangles indicate the sign of the vector chirality.  (b)~The structure of jarosite is shown as a ball-and-stick model, highlighting the Fe$^{III}$$_3$($\mu$-OH)$_3$ bond connectivity.} \label{fig1}
\end{figure}

Frustrated spin systems have received considerable attention from both theorists and experimentalists due to their unconventional ground states and spin dynamics such as spin liquids,\cite{5,6,8,9} spin nematic,\cite{7} and spin ice.\cite{ramirez:1,schiffer:2} Antiferromagnetically coupled spins located on the vertices of corner-sharing triangles that make up the kagome lattice (Figure~\ref{fig1}(a)) are highly sought after as being one of the most highly frustrated spin systems on a two-dimensional (2D) lattice.\cite{1,2} Among other realizations of the kagome lattice antiferromagnet such as SrCr$_9$Ga$_3$O$_{19}$,\cite{ramirez:2070,broholm_SCGO} volborthite,\cite{hiroi,Yoshida:2009p7543,Bert:2005p1111} herbertsmithite,\cite{shores:053891,helton:107204,Helton:2010p10424,Mendels:2007p335,Lee:2007p1219} and Rb$_2$Cu$_3$SnF$_{12}$,\cite{Morita:2008p2717,Ono:2009p4220,Matan:2010p10691}, jarosite\cite{wills,inami,Nishiyama:2003p1080,shl_cr, 18,20,Coomer:2006p70,Bisson:2008p5528} is considered one of the most ideal systems for the following reasons.  First, it consists of single and undistorted kagome planes with no structural phase transition down to the base temperature. Second, a stoichiometrically pure, single-crystal sample can be synthesized.\cite{grohol_chiral}  

The general chemical formula of materials in this family is $AM_3$(OH)$_6$(SO$_4$)$_2$, where $A$ is a monovalent or bivalent cation such as Na$^+$, K$^+$, Rb$^+$, NH$_{4}^{+}$, Ag$^+$, Tl$^+$, H$_3$O$^{+}$, or $\frac{1}{2}$Pb$^{2+}$, and $M$ is a trivalent cation such as Fe$^{3+}$, Cr$^{3+}$, or V$^{3+}$. For jarosite, the magnetic $S=5/2$ Fe$^{3+}$ ions, which lie at the center of a tilted octahedral cage formed by six oxygen atoms, reside at the corners of the triangles of the kagome lattice. The kagome planes, which are separated by nonmagnetic cations $A$ and sulfate groups, stack up along the $c$-direction with weak interlayer coupling, attesting to the two-dimensionality of this system.

Even though jarosite has long been regarded as a prime model for studying 2D spin frustration, until recently the inability to synthesize magnetically pure samples had clouded efforts to fully understand the precise magnetic behavior of this system. We have developed kinetic control of the sample preparation process through oxidation-reduction chemistry, which leads to full occupancy of the magnetic sites.\cite{18,19,20} This technique affords materials that display consistent magnetic properties. Most notably, we observe a prominent antiferromagnetic ordering temperature $T_N$ narrowly ranging from 56-65 K for jarosites with different interplanar cations.\cite{18,20} In addition, the nearest-neighbor exchange interaction is consistent for all jarosites, indicated by similar values of the Curie-Weiss temperature of $\sim-800$ K. This similarity in the Curie-Weiss temperature is not surprising, as the basic magnetic unit, the Fe$^{III}$$_3$($\mu$-OH)$_3$ triangle, is structurally homologous in all jarosites.

Due to frustration and low dimensionality, ground states of a Heisenberg kagome lattice antiferromagnet are infinitely degenerate. Hence, no three-dimensional (3D) magnetic  order is expected at any non-zero temperature.  However, in jarosite, anisotropic terms and small interlayer coupling in the spin Hamiltonian are present, and the 3D magnetic long-range order (LRO) has been observed at low temperature $T<T_N$. Powder neutron diffraction measurements on ${\rm KFe_3(OH)_6(SO_4)_2}$ show that the spins order in a coplanar $q=0$ structure (Figure~\ref{fig1}(a)) below $T_N \approx 65$~K.\cite{inami,Wills2} In the $q=0$ structure, spins on each triangle form the $120^\circ$ arrangement within the kagome plane, and point either toward (all-in) or away from (all-out) the center of the triangle.

\begin{figure}[t]
\centering
\includegraphics[width=2.8in]{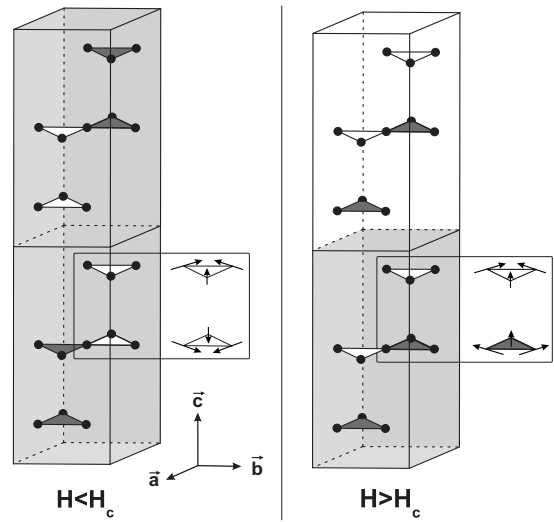}
\caption{Spin re-orientations are shown on the second, fourth and sixth layers (from the bottom) when $H>H_c$. White triangles indicate the all-in arrangement and dark triangles indicate the all-out arrangement. The shaded area shows the volume of the magnetic unit cell.} \label{fig2}
\end{figure}

The ordered spins can be decomposed into three sublattices, with the spins on each triangle labeled \textbf{S$_1$}, \textbf{S$_2$} and \textbf{S$_3$} as shown in Figure~\ref{fig1}(a).  The vector chirality for each triangle can be defined as:
\begin{equation}
\textbf{K}_v =
\frac{2}{3\sqrt{3}}(\textbf{\^S$_1$}\times\textbf{\^S$_2$}+\textbf{\^S$_2$}\times\textbf{\^S$_3$}+\textbf{\^S$_3$}\times\textbf{\^S$_1$}).\label{kv}
\end{equation}
For the coplanar arrangement, this vector is parallel to the $c$-axis with amplitude of +1 or -1.  A triangle has positive (negative) vector chirality if the spins rotate clockwise (counterclockwise) as one traverses around the vertices of the triangle clockwise.  Figure~\ref{fig1}(a) shows the $q=0$ structure with positive vector chirality.  Figure~\ref{fig2} shows the stacking of the kagome layers along the $c$-direction, where white triangles represent the all-in arrangement and black triangles represent the all-out arrangement.  The scalar chirality for each triangle can also be defined as:
\begin{equation}
K_s=\textbf{S}_1\cdot\left(\textbf{S}_2\times \textbf{S}_3\right).\label{ks}
\end{equation}
The scalar chirality on a single plane is non-zero if spins cant out of the plane.  We have previously reported the field-induced spin-canting transition that corresponds to a non-trivial change of the scalar chirality from the net $K_s=0$ to $K_s\neq0$ state.\cite{grohol_chiral}

The 3D magnetic LRO in jarosite is a result of ferromagnetic interlayer coupling and spin anisotropy arising from the Dzyaloshinskii-Moriya (DM) interaction.\cite{matan, grohol_chiral, yildirim} The DM interaction, which is a perturbation on the Heisenberg spin Hamiltonian, is present if there is no inversion center between magnetic sites, and its strength is linearly proportional to the spin-orbit coupling.\cite{37,dzyaloshinskii,moriya} It has been experimentally observed in spin frustrated perovskite cuprates,\cite{23,24,25,26,27,28,29,30} the pyrochlore antiferromagnet Cu$_4$O$_3$,\cite{31} and molecule-based magnets.\cite{32,33,34,35,36} The microscopic calculations of the DM interaction on the kagome lattice have been carried out by Elhajal \textit{et al.}\cite{37} by applying Moriya's formulation.\cite{moriya} To first approximation, the spin Hamiltonian is given by:
\begin{equation}
{\cal H} = \sum_{nn} \left[ J_1 \textbf{S}_i \cdot \textbf{S}_j + \textbf{D}_{ij}\cdot \textbf{S}_i \times \textbf{S}_j \right],\label{model_H}
\end{equation}
where $\Sigma_{nn}$ indicates summation over pairs of nearest neighbors, $J_1$ is the nearest-neighbor exchange interaction, and \textbf{\textit{D}}$_{ij}$ is the DM vector defined in a local axis.\cite{yildirim} Our spin wave measurements on a single-crystal sample provide all relevant spin Hamiltonian parameters for this Heisenberg spin model.\cite{matan, yildirim, matan_pol}

\begin{figure}[t]
\centering
\includegraphics[width=2.8in]{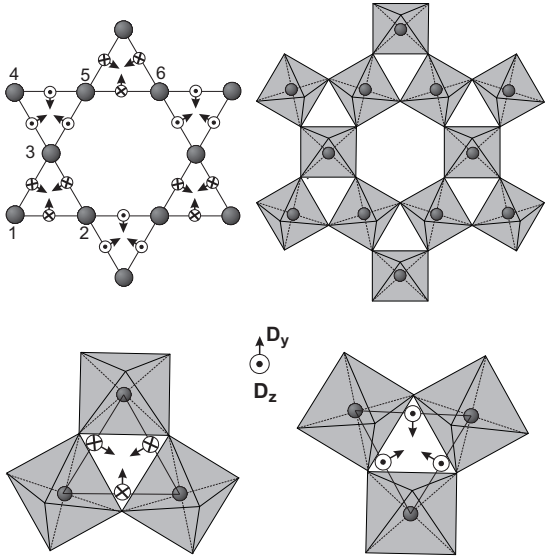}
\vspace{0mm} \caption{DM vectors on the kagome lattice.  The tilt of the octahedra is shown in the lower diagram.  The DM vector lies within the mirror plane bisecting the bond connecting two magnetic sites.  In this case, it is assumed that the outward (inward) tilt of an octahedron corresponds to the DM vector pointing into (out-of) the page.}\label{fig3}
\end{figure}

Figure~\ref{fig3} shows the DM vector, which lies on a mirror plane bisecting the bond connecting two magnetic sites, with respect to the tilt of the FeO$_6$ octahedra.  If one octahedron tilts outward, the octahedron that lies right on top or below will tilt inward.  Therefore, the direction of the out-of-plane component of the DM vector changes from out-of to into the page or vice versa when one moves up or down one layer. On the other hand, the in-plane component always points toward the center of the triangle.  Unfortunately, these symmetry considerations alone do not determine a precise direction or value of the DM vector, which depends on microscopic details, and can only be calculated by a microscopic theory.  However, once the DM vector is calculated for one bond, all other DM vectors can be determined using symmetries of the crystal. 

The DM interaction can induce a magnetically ordered state at a non-zero temperature by lifting the macroscopic ground-state degeneracy, and determines the ground-state spin structure.\cite{37,38} In jarosite, the tilt of the octahedra is about 20$^\circ$ with respect to the crystallographic $c$-axis,\cite{18,20} resulting in both out-of-plane $D_z$ and in-plane $D_p$ components of the DM vector. Elhajal \emph{et al.} show that the DM interaction can give rise to LRO in the kagome lattice, and induce spins to cant out of the kagome planes to form an ``umbrella'' structure. $D_z$ confines the spins to lie within the kagome planes, and hence effectively takes a role of an easy-plane anisotropy.  Furthermore, it breaks the symmetry between positive and negative vector chirality, leading to the $q=0$ spin structure.  On the other hand, $D_p$, which takes a role of an easy-axis anisotropy, breaks the rotational symmetry about the $c$-axis and gives rise to the all-in-all-out spin arrangement. It also forces the spins to cant out of the kagome planes to form the umbrella structure of ferromagnetically aligned canted moments giving each layer a non-zero magnetic moment.\cite{37,38,39} However, in the absence of an applied field, the interlayer coupling prompts the canted moments on neighboring planes to align antiferromagnetically along the $c$-axis (Figure~\ref{fig2}), yielding a zero net magnetic moment.

In this article, we report the results of specific heat, magnetization, and neutron scattering measurements on two jarosites, ${\rm KFe_3(OH)_6(SO_4)_2}$ (K jarosite) and ${\rm AgFe_3(OH)_6(SO_4)_2}$ (Ag jarosite). In Section II, we discuss the specific heat and magnetization measurements, and study the field-induced transition between states with different spin textures. In Section III, we investigate spin re-orientation at the transition using neutron scattering. The result provides the first direct evidence of the long-expected spin rotation. In Section IV, the phase diagrams of Ag and K jarosites are discussed. 

\begin{figure}[t]
\centering
\includegraphics[width=2.8in]{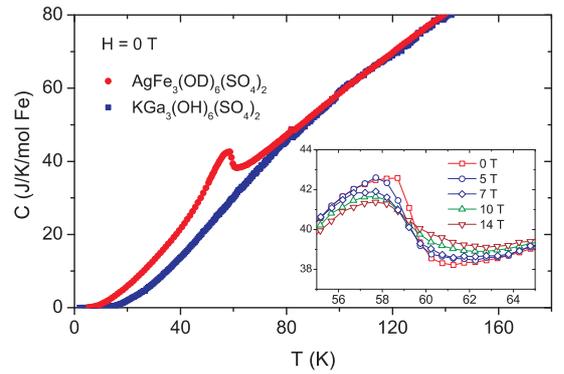}
\caption{(Color online) Specific heat of polycrystalline samples of deuterated Ag jarosite (red circles), and of non-magnetic isostructural compound KGa$_3$(OH)$_6$(SO$_4$)$_2$ (blue squares) measured at $H=0$ T. The KGa$_3$(OH)$_6$(SO$_4$)$_2$ data are scaled to match the Ag jarosite data at high temperature and used to estimate lattice contribution. The inset shows the specific heat of Ag jarosite at different magnetic fields. Uncertainties where indicated in this article are statistical in origin and represent one standard deviation.} \label{fig4}
\end{figure}

\section{Specific heat and magnetization measurements}
Details of the synthesis and X-ray crystal structure refinement of Ag jarosite have been previously reported in Ref.~\onlinecite{20}, whereas those of K jarosite have been reported in Refs.~\onlinecite{18} and~\onlinecite{grohol_chiral}. The specific heat $C$ of a deuterated polycrystalline sample of Ag jarosite was measured over a 5-150 K temperature range at field strengths varying from 0-14 T (Figure~\ref{fig4}). At zero field,  a cusp  at $T_N\approx$ 59 K indicates the transition to the 3D magnetic LRO. When the field is aligned along the $c$-axis, the cusp broadens with increasing field (the inset of Figure~\ref{fig4}). The entropy associated with the magnetic order, which is obtained from integrating the magnetic part of $C/T$ (after correcting for lattice contribution which was estimated from the non-magnetic isostructural compound KGa$_3$(OH)$_6$(SO$_4$)$_2$) over the temperature range from 5 K to 150 K, represents about 70\% of the $R\ln6$ (where $R$ is the molar gas constant) total entropy expected from the spin-5/2 system, compared with $\sim50\%$ for K jarosite.\cite{grohol_chiral} This result suggests that short-range correlations have already formed at temperature higher than 150 K.  

\begin{figure}[t]
\centering
\includegraphics[width=2.8in]{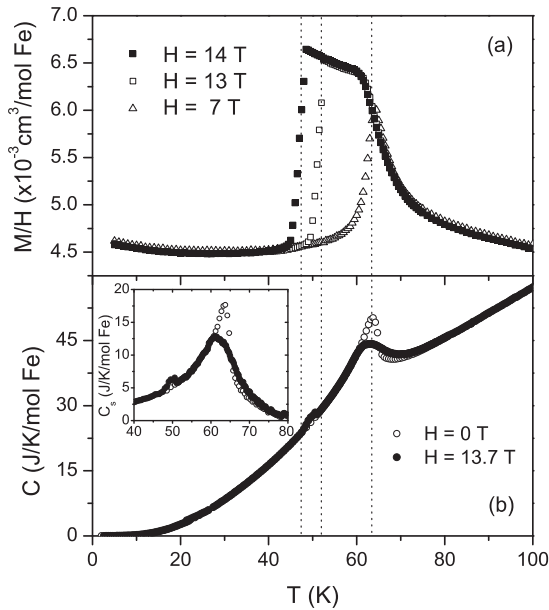}
\vspace{0mm} \caption{Magnetic susceptibility and specific heat were measured on single-crystal K jarosite with magnetic fields up to 14 T. A field was applied parallel to the $c$-axis.  (a)~$M/H$ as a function of temperature measured at $H=$7 T (closed squares), 13 T (open squares), and 14 T (open triangles). (b)~Specific heat as a function of temperature measured at $H=0$ T (open circles) and $13.7$ T (closed circles). The inset shows the spin-only specific heat.}\label{fig5}
\end{figure}

The specific heat of single-crystal K jarosite measured with the field of 13.7 T along the $c$-axis shows two anomalies, one at 64 K and the other at 50 K (Figure~\ref{fig5}).  The former coincides with the transition to the ordered state in zero field. The sharp peak is, however, replaced by a broad cusp in the presence of  the magnetic field. The latter occurs at the same temperature where the sharp drop in the magnetization is observed, corresponding to the spin re-orientation transition between the ferromagnetic and antiferromagnetic states.  This anomaly is only observed in the single-crystal sample, but not in the polycrystalline sample due to powder average and its small contribution to the overall specific heat.

Magnetization measurements on polycrystalline samples of non-deuterated and deuterated Ag jarosite were performed over a 5-300 K temperature range at field strengths varying from 0-14 T. All raw magnetization data were corrected for diamagnetism due to core electrons, and paramagnetic contributions due to free spins or impurities. To extract the paramagnetic contributions, the low field region of the magnetization $M(H)$ for $H<9.5$ T at $T=5$ K was fit to $M(H)=P_1B_J(H/T)+P_2H+P_3$, where $P_n$ are empirical pre-factors and $B_J(x)$ is the Brillouin function with $J = 5/2$. The average value of $P_1$ upon increasing and decreasing fields at 5 K is about $0.005~\mu_B$, which indicates that about $0.1\%$ of the Fe$^{3+}$ spins are non-interacting. This result suggests that all magnetic sites are occupied, and Ag jarosite has virtually no non-interacting spins in the ordered state.

\begin{figure}[t]
\centering
\includegraphics[width=2.8in]{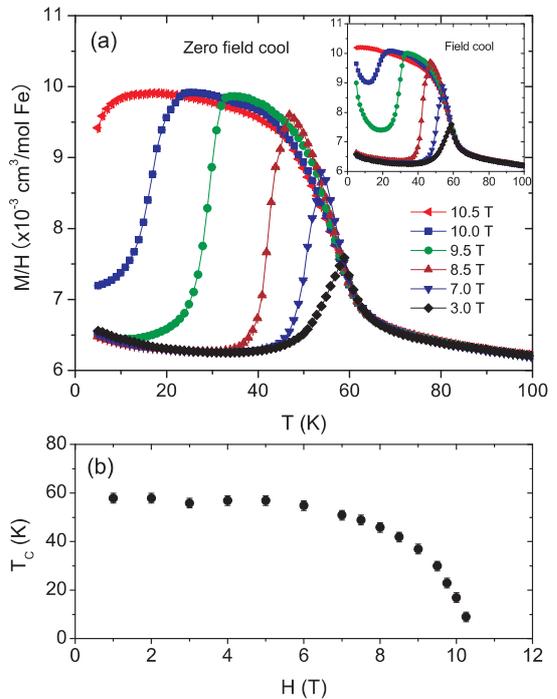}
\caption{(Color online) (a)~Zero field-cool and field-cool (inset) magnetic susceptibility of deuterated Ag jarosite. (b)~shows $T_c$ as a function of field.} \label{fig6}
\end{figure}

The magnetic susceptibility ($\chi=M/H$) measurements on non-deuterated samples at low field of 100 Oe (not shown) show LRO below the N\'eel ordering temperature $T_N$ of 59.6 K.\cite{20} At high temperature (T$>150$ K), $\chi$ follows the Curie-Weiss law $\chi=C'/(T-\Theta_{CW}')$. A fit to this equation yields $\Theta_{CW}'=-800(50)$ K and $C'=5.4(5)$ cm$^3$ K/mol Fe. The ordering temperature $T_N$ is greatly reduced from the mean field value $|\Theta_{cw}'|$ due to geometric frustration, suggested by a large empirical frustration parameter,\cite{1} $f = |\Theta_{cw}'|/T_N\approx13$. Therefore, the mean field theory cannot be applied to determine the effective moment $\mu_{eff}$, and the nearest neighbor exchange coupling $J_1$. Harris \emph{et al.}\cite{harris} introduce the corrections to the mean field theory to account for geometric frustration using the high-temperature series analysis for the kagome lattice, which gives $\Theta_{CW}'=\frac{3}{2}\Theta_{CW}$ and $C'=\frac{9}{8}C$ where $\Theta_{CW}$ and $C$ are the mean field Curie-Weiss temperature and constant, respectively. Using these corrections, we obtain the effective moment $\mu_{eff}=6.2(6)~\mu_{B}$ and the nearest neighbor exchange coupling $J_1=3.9(2)$ meV. For comparison, the values of the nearest neighbor coupling and the effective moment for K jarosite deduced from the Curie-Weiss fit using Harris' corrections are $J_1=3.9(2)$ meV and $\mu_{eff}=6.3(2)~\mu_{B}$,\cite{grohol_chiral} respectively. We note the similarity of the values of $J_1$ and $\mu_{eff}$ for K jarosite and Ag jarosite due to the structural homology of the fundamental interacting unit, the Fe$^{III}$$_3$($\mu$-OH)$_3$ triangle. Goodenough-Kanamori rules governing the frontier orbitals of the hydroxide-bridged iron trimer predict this antiferromagnetic exchange.\cite{43,44} The singly occupied $d_{x^2-y^2}$ orbitals of high spin Fe$^{3+}$ overlap with the filled $sp^3$ hybrid orbitals of the hydroxide, giving a superexchange pathway of $\sigma$-symmetry.

Figure~\ref{fig6}(a) shows the magnetic susceptibility of deuterated Ag jarosite after subtracting the paramagnetic part.  All measurements were performed on warming at fields between 3 T and 14 T.  At high fields, the peak at $T\approx60$ K becomes a broad plateau.  The plateau is getting broader as the field increases, and extends to the base temperature of 5 K for $H>10.5$ T.  The zero-field-cool and field-cool data show the same temperature dependence except at low temperature where the field-cool data show a rise in the susceptibility, which we believe is due to trapped ferromagnetic domains of the polycrystalline sample.  The susceptibility of the non-deuterated sample (not shown) shows the the same temperature and field dependences but the saturated value at high fields ($\chi_{max}\approx9\times10^{-3}$ cm$^{3}$/mol Fe) is about 10\% smaller than that of the deuterated sample.

\begin{figure}[t]
\centering
\includegraphics[width=2.8in]{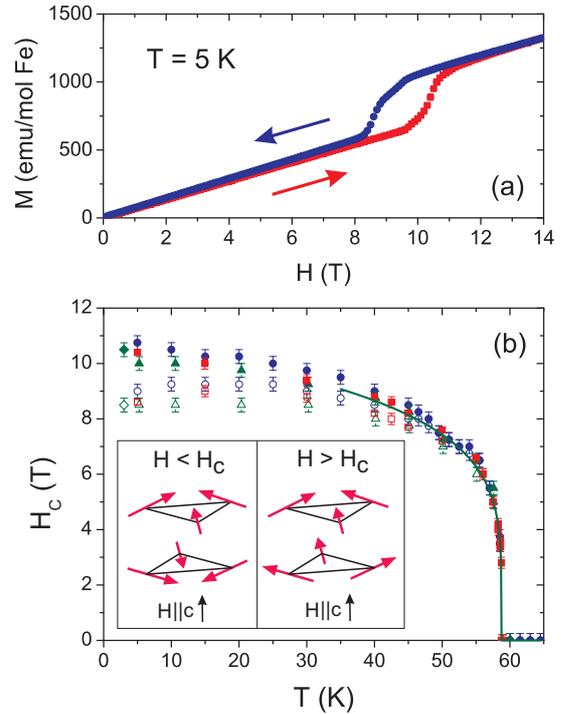}
\caption{(Color online) (a)~Magnetization as a function of field at $T=5$ K for deuterated Ag jarosite shows hysteresis as the field is swept up and down (1 emu = $10^{-3}$ A$\cdot$m$^2$). (b)~Temperature dependences of the critical field $H_c$  between 5 K and 60 K are obtained from magnetization measurements on non-deuterated Ag jarosite (blue circles) and deuterated Ag jarosite (red squares). Neutron scattering results on deuterated Ag jarosite measured at BENSC and NCNR are denoted by green diamonds and green triangles, respectively. A green curve corresponds to a power-law fit to $a|T-T_N|^\beta$ between $35<T<65$ K with exponent $\beta=0.20(1)$ and $T_N=58.8(3)$ K. Open and closed symbols denote decreasing and increasing field sweeps, respectively. The inset shows the spin re-orientation transition of the first layer from the all-in arrangement for $H<H_c$ to the all-out arrangement for $H>H_c$.} \label{fig7}
\end{figure}

Figure~\ref{fig6}(a) also shows sharp decreases in the susceptibility in the ordered state. By taking the derivative of the susceptibility with respect to temperature, we can define $T_c$ at a peak in the derivatives. $T_c$, which is field-dependent, indicates the transition between the ferromagnetic and antiferromagnetic states. Figure~\ref{fig6}(b) shows $T_c$ as a function of field for deuterated Ag jarosite. For $T<T_c(H)$, the system is in the magnetically ordered state with no net ferromagnetic moment. In this state, the net scalar chirality is zero.\cite{grohol_chiral}  For $T_c(H)<T<T_N$, the system is also in the magnetically ordered state, but the net scalar chirality is now non-zero and the net ferromagnetic moment is present. From the magnetization and specific heat measurements, it is difficult to precisely define the ordering temperature $T_N$ at high fields due to the lack of well-defined peaks.  However, $T_N$ can be readily measured by neutron scattering, which will be discussed Section III.

\subsection{Critical field, $H_c$, and canted moment, $\Delta M$}

High field magnetization measurements on polycrystalline samples of Ag jarosite were performed up to a maximal field of $14~$T; the similar measurements on a single-crystal sample of K jarosite were previously reported in Ref.~\onlinecite{grohol_chiral}. Figure~\ref{fig7}(a) shows the magnetization $M$ at $T=5~$K as a function of field after subtracting the paramagnetic part. $M$ increases linearly with increasing applied field up to $9.75~$T, which is followed by a sharp increase between $9.75 -11.5~$T.  The linear response is then recovered for $H > 11.5~$T. Continuing the measurement with decreasing applied field, $M$ decreases linearly to $10.5~$T, then drops sharply between $8.75-10.5~$T.  The linear behavior then resumes for $H < 8.75~$T. The hysteresis in the bulk magnetization signifies ferromagnetic ordering, and is indicative of a first-order transition between the antiferromagnetic and ferromagnetic states. The anisotropic crystal field of Fe$^{3+}$ ions in a distorted octahedral geometry yields the hysteresis in the $M$-$H$ curve with a coercive field of approximately 2~T at $T=5$~K. The critical field $H_c$, at which the sudden change in magnetization is observed, is defined as a maximum of $(dM/dH)|_T$. Upon increasing (decreasing) field, $H_c = 10.75~$T ($8.75~$T) at $T=5$ K. The same measurements were conducted at different temperatures to obtain $H_c(T)$. The results are shown in Figure~\ref{fig7}(b). The line corresponds to a fit to the power law $H_c(T)\propto|T-T_N|^\beta$ for a temperature range between 35 K and 65 K giving exponent $\beta=0.20(1)$ and $T_N=58.8(3)$ K compared with $\beta=0.25$ and $T_N=64.7$ K for K jarosite.\cite{grohol_chiral}

Additional high-field magnetization measurements on a single-crystal sample of K jarosite were performed using a nondestructive pulsed magnet, which can generate pulsed magnetic fields up to 55 T, at the International MegaGauss Science Laboratory, Institute for Solid State Physics, The University of Tokyo.  These pulsed-field measurements are essential to determine the canted moment and critical field in K jarosite at low temperature, since the critical field below 50 K exceeds 14~T, the maximal field of our previous measurements.\cite{grohol_chiral}  Magnetizations were measured by induction using a coaxial pick-up coil.  A single-crystal sample of K jarosite was aligned so that the magnetic field was applied perpendicular to the kagome plane.  The sample was then cooled in a liquid $^4$He cryostat down to 4.2 K.  Raw data were normalized to magnetizations reported in Ref.~\onlinecite{grohol_chiral}.

\begin{figure}[t]
\centering
\includegraphics[width=3.6in]{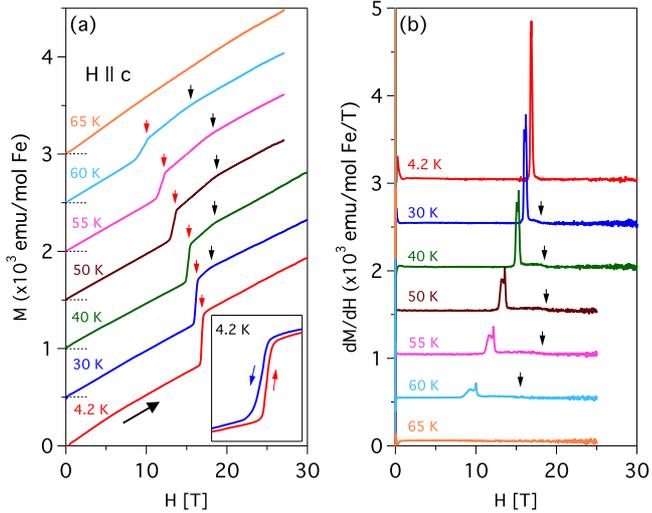}
\caption{(Color online) High-field magnetization measurements using a pulsed magnet on single-crystal K-jarosite. (a) Magnetization curves at several temperatures show a sharp rise at $H_c$ indicated by red arrows and a kink indicated by black arrows. (b) dM/dH curves show a clear peak, which is used to define $H_c$.  The data in (a) and (b) are offset for clarity.} \label{fig8}
\end{figure}

We also observed the hysteretic behavior, indicative of the first-order transition, in K jarosite at low temperature as shown in the inset of Figure~\ref{fig8}.  The coercive field in K jarosite is, however, much smaller than that in Ag jarosite.  The hysteresis becomes smaller as temperature increases, and is undetectable above 50 K.  Nevertheless, we cannot rule out the possibility that the hysteresis exists up to $T_N$. In addition to the transition at $H_c$, the magnetization curves show an additional kink at $H_k>H_c$ denoted by black arrows in Figures~\ref{fig8}(a) and~\ref{fig8}(b).  As temperature decreases, $H_k$ and $H_c$ converge, and the high-slope region between $H_k$ and $H_c$ disappears at the base temperature.  This kink was not clearly observable in the polycrystalline sample of Ag jarosite.  It is possible that the prominent first-order transition in Ag jarosite suppresses this intermediate phase.  A similar kink in the $M$-$H$ curve was also observed in ferromagnetic perovskites, which show a change from first- to second-order magnetic phase transitions as a function of doping.\cite{Mira:1999p9074}

\begin{figure}[t]
\centering
\includegraphics[width=2.8in]{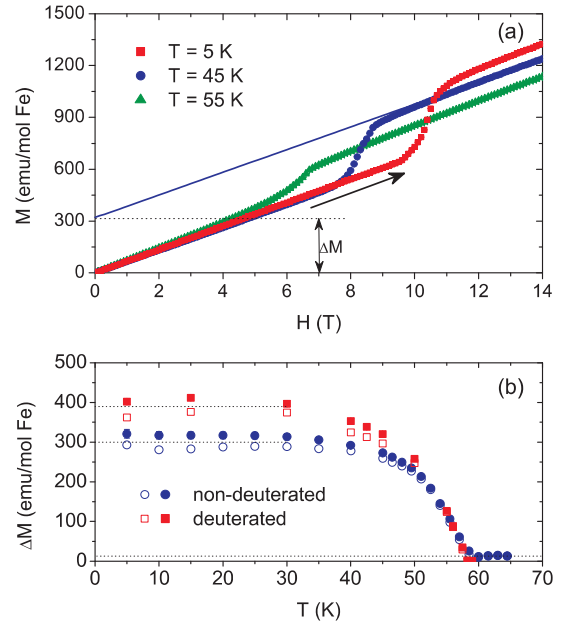}
\caption{(Color online) (a)~Magnetization as a function of field at different temperatures for deuterated Ag jarosite. The $H>H_c$ range is fit to a linear function with the same slope as that of the $H<H_c$ range. The canted moment $\Delta M$ is defined as a $y$-intercept. (b)~shows temperature-dependence of $\Delta M$ for non-deuterated Ag jarosite (blue circles) and deuterated Ag jarosite (red squares). $\Delta M$ saturates at $300(30)~$emu/mol Fe ($390(30)~$emu/mol Fe) for the non-deuterated (deuterated) sample. Open and closed symbols denote decreasing and increasing field sweeps, respectively.} \label{fig9}
\end{figure}

To extract a canted moment $\Delta M$ from the data, we fit the high field data ($H>H_c$) of the $M$-$H$ curve to a linear function with a fixed slope and extrapolated that line to intercept the $y$-axis to obtain $\Delta M$ (Figure~\ref{fig9}(a)). The slope was obtained from fitting low field data ($H<H_c$) to a linear function fixing a constant term to zero.  The hysteresis in $\Delta M$ persists up to $T\approx50$ K, which is in agreement with the hysteresis observed in the critical field (Figure~\ref{fig7}(b)).  Below $T\approx50$ K, the canted moments measured on increasing field are slightly larger than those measured on decreasing field.   Figure~\ref{fig9}(b) also shows the difference of $\Delta M$ between deuterated and non-deuterated samples.  This difference could be due to a difference in the in-plane component of the DM interaction $D_p$ arising from the replacement of hydrogen by deuterium.

The presence of the canted moment allows us to investigate the DM interaction and the interlayer coupling directly. The sudden change in magnetization is a result of the difference in the magnetic moment along the $c$-axis caused by 180$^\circ$ rotation of spins on the alternate layers that were previously oppositely canted due to the ferromagnetic interlayer coupling as shown in the diagram in Figure~\ref{fig7}(b). At the critical field, the magnetic field energy overcomes the ferromagnetic interlayer coupling, and forces the spins on the alternate layers to rotate 180$^\circ$ causing the canted moments to align. 

\subsection{Canting Angle $\eta$, DM Parameters $D_p$ and $D_z$, and interlayer coupling $J_\bot$}

The canting angle, DM parameters, and interlayer coupling can be calculated from the magnetization results ($H_c$ and $\Delta M$). $\Delta M$ saturates for temperature below 30 K, and goes to zero as temperature approaches $T_N$. The dashed lines in Figure~\ref{fig9}(b) show that $\Delta M$ saturates at $\Delta M(0)=300(30)~$emu/mol Fe $=0.054(5)~\mu_B$
($390(30)~$emu/mol Fe $=0.070(5)~\mu_B$) for non-deuterated (deuterated) Ag jarosite. However, since the data were measured on polycrystalline samples, $\Delta M$ is multiplied by three to take into account the powder average. Given the expected moment $M=5~\mu_B$ for $S=5/2$, the canting angle at zero temperature $\eta$ is defined as $\eta=\sin^{-1}\left(\frac{\Delta M(0)}{M}\right)$, which gives $\eta=1.8(2)^\circ$ ($\eta=2.4(2)^\circ$) for the non-deuterated (deuterated) sample. For K jarosite, $\Delta M(0)=650(80)$ emu/mol Fe and hence $\eta=1.3(2)^\circ$, which is lower than the value of 1.9(2)$^\circ$ obtained from the spin wave measurements.\cite{matan} This difference could be due to thermal or quantum fluctuations, which reduce the staggered moment.  The similar measurements of the spin canting in La$_2$CuO$_4$, a parent compound of high-$T_c$ superconductors, yields a much smaller value of the canting angle ($\eta=0.17^\circ$).\cite{thio}

\begin{table*}
\caption{Critical fields, canted moments at zero temperature, canting angles, and spin Hamiltonian parameters for K jarosite and Ag jarosite calculated from magnetization and neutron scattering data.}
\begin{center}
\begin{tabular}{| l | c | c | c | c | c | c | c |}
\hline \hline
Compound & H$_\text{C}$(0) (T) &  $\Delta$M(0) ($\mu_B$) & $\eta$ ($^\circ$)&$|$D$_p|$ (meV)& D$_z$ (meV)& J$_c$ (meV) &$\hbar\omega_0$ (meV)\\
\hline
AgFe$_3$(O$X$)$_6$(SO$_4$)$_2$ &  & & & & & & \\
~~I.~~$X=$ H & 9.5(1.0)& 0.16(2) & 1.8(2) & 0.18(3) & $-0.17(3)$ & $-0.014(3)$ & 6.2(4)\\
~~II.~$X=$ D & 9.5(1.0) & 0.21(2)& 2.4(2)&0.24(3)&$-0.17(3)$& $-0.018(4)$& 6.2(4)\\
\hline
KFe$_3$(OH)$_6$(SO$_4$)$_2$ &  & & & & & & \\
~~I.~~Neutron scattering\cite{matan} & $-$ & $-$ & 1.9(2)&0.197(2)&$-0.196(4)$&$-0.027(4)$&6.7(1)\\
~~II.~Magnetization & 16.8(5) & 0.12(2)& 1.3(2)& $-$ & $-$ &$-0.019(4)$&$-$\\
\hline \hline
\end{tabular}
\end{center}
\label{table1}
\end{table*}

Since the spin canting is a result of the DM interaction, one can write down the expression for $\eta$ in terms of the DM parameters. Elhajal \emph{et al.} derived the following relation between $\eta$ and the DM parameters $D_p$ and $D_z$:\cite{37}
\begin{equation}
\eta=\left|\frac{1}{2}\tan^{-1}\left[\frac{-2D_p}{\sqrt{3}J_1-D_z}\right]\right|.
\label{canting1}
\end{equation}
We note that signs of $D_p$ and $D_z$ are reversed and different from Ref.~\onlinecite{37} to be consistent with the DM vector defined in this article and in Ref.~\onlinecite{yildirim}. The absolute-value indicates that the sign of the canting angle with respect to the in-plane order cannot be uniquely determined from the DM vector alone. In fact, for a given value of $D_p$ with $D_z<0$, there exist two lowest-energy degenerate states with opposite canting and different in-plane orders corresponding to the $K_s>0$ and $K_s<0$ states.\cite{yildirim} At zero field, this chiral symmetry is spontaneously broken at $T_N$, and the interlayer coupling selects the spin structure on neighboring planes, giving rise to the 3D magnetic LRO. Equivalently, applying a magnetic field along the $c$-axis will also lift this degeneracy.

\begin{figure}[t]
\centering
\includegraphics[width=2.8in]{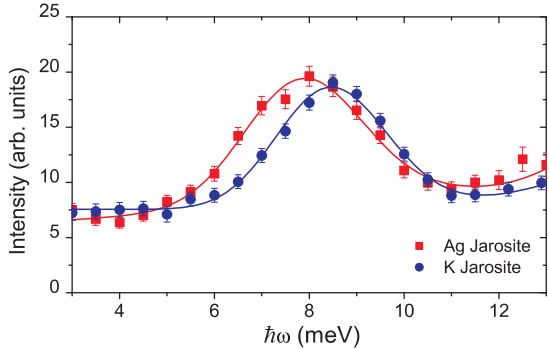}
\caption{(Color online) Inelastic neutron scattering measurements were performed using the triple-axis spectrometers at NCNR. The constant-Q scan at $|Q|=1.7$ \AA~ on deuterated Ag jarosite (red squares) was measured at BT7, and the constant-Q scan at $|Q|=1.5$ \AA~ on deuterated K jarosite (blue circles) was measured at BT9. The scans show powder-average spin-wave excitations at energy transfer of 7.9(4) meV and 8.5(3) meV for Ag jarosite and K jarosite, respectively. The K jarosite data were normalized so that they lie on top of the Ag jarosite data. Solid lines show the fits to the spin wave excitations described in Ref.~\onlinecite{matan}, convoluted with the instrumental resolution function.} \label{fig10}
\end{figure}

Yildirim and Harris have also derived a similar expression for the canting angle induced by the DM interaction:\cite{yildirim}
\begin{equation}
\eta=\left|\frac{1}{2}\sin^{-1}\left[\frac{-2D_p}{\sqrt{3}J-D_z}\right]\right|.
\label{canting2}
\end{equation}
where $J=J_1+J_2$. Both Eq.~\ref{canting1} and Eq.~\ref{canting2} approach the same limiting value for a small canting angle. In order to determine both in-plane $D_p$ and out-of plane $D_z$ components of the DM interaction, we also need the value of the energy gap of the lifted zero energy mode, and therefore rely on previous single-crystal neutron scattering measurements on K jarosite.\cite{matan} The spin wave calculations by Yildirim and Harris give the following expression for the two energy gaps around 7 meV at $\Gamma$ point:\cite{yildirim}
\begin{equation}
\hbar\omega_0=S\left(\sqrt{3D^2_p+18D_z^2-6\sqrt{3}JD_z}\pm\frac{2D_zD_p}{J}\right).\label{zero_E}
\end{equation}
The energy splitting of these two gaps is about 0.1 meV, where the energy of the lifted zero energy mode has a smaller value.  This value is independent of the sign of $D_p$ due to the $\pm$ sign in front of the last term. $|D_p|$ and $D_z$ can be calculated by solving Eqs.~\ref{canting2} and \ref{zero_E} providing $\eta$ and $\hbar\omega_0$. Given $\hbar\omega_0=6.7(1)$ meV from the single-crystal neutron scattering measurements\cite{matan} and using the data shown in Figure~\ref{fig10}, we estimated the energy gap for Ag jarosite to be 6.2(4) meV.  Therefore, after solving Eqs.~\ref{canting2} and \ref{zero_E}, we obtain $|D_p|=0.18(3)$ meV (0.24(3) meV), and $D_z=-0.17(3)$ meV ($-0.17(3)$ meV) for non-deuterated (deuterated) Ag jarosite (see Table~\ref{table1}). Here, we assume that $J$ is the same for Ag jarosite and K jarosite as evidenced by the the same Curie-Weiss temperature.

In addition, the knowledge of $\Delta M(0)$ and $H_c(0)$ from the magnetization measurements allows us to calculate the interlayer coupling $J_\bot$. Similar to $\Delta M$, $H_c$ appears to saturate at low temperature to the average value of $H_c(0)=9.5(1.0)~$T. As temperature increases, $H_c$ monotonically decreases and goes to zero as temperature approaches $T_N$. The field-induced transition is a result of the competition between the ferromagnetic interlayer coupling and the Zeeman energy. Using the equation $J_\bot=-\Delta M(0)\cdot H_c(0)/S^2$,\cite{thio,grohol_chiral} we obtain $J_\bot=-0.014(3)$ meV ($-0.018(4)$ meV) for the non-deuterated (deuterated) Ag jarosite.   The minus sign indicates that the interlayer coupling is ferromagnetic.  For K jarosite, using the canting angle of 1.9(2)$^\circ$ that gives $\Delta M(0)=0.17(2) \mu_B$ from the spin wave measurements, or $\Delta M(0)=0.12(2) \mu_B$ from the pulsed-filed magnetization measurements, we obtain $J_\bot=-0.027(4)$ meV or -0.019(4) meV, respectively (see Table ~\ref{table1}). The magnitude of $J_\bot$ is hundreds times smaller than the magnitude of $J_1$, which attests to the two dimensionality of this spin system.

\section{Neutron scattering measurements}
Inelastic neutron scattering measurements in zero field were performed using the thermal triple-axis spectrometers BT7 and BT9 at the NIST Center for Neutron Research (NCNR) with the final energy fixed at 14.7 meV to measure spin wave excitations on deuterated polycrystalline Ag and K jarosites, respectively. The horizontal collimations of open$-50^\prime-$S$-60^\prime-120^\prime$ and $40^\prime-56^\prime-$S$-52^\prime-$open were employed at BT7 and BT9, respectively. Pyrolytic graphite (PG) crystals were utilized to monochromate and analyze the incident and scattered neutron beams.  PG filters were placed in the scattered beam to reduce higher-order contamination. The samples were cooled using a closed cycle $^4$He cryostat to $T=5.3$ K at BT7 and to $T=14$ K at BT9. The energy scans in Figure~\ref{fig10} at $|Q|=1.7$ \AA~ for Ag jarosite (red squares) and $|Q|=1.5$ \AA~ for K jarosite (blue circles) show powder-average spin wave excitations of the lifted zero-energy mode,\cite{matan} which is weakly dispersive, at energy transfer of 7.9(4) meV and 8.5(3) meV, respectively.

\begin{figure}[t]
\centering
\includegraphics[width=2.8in]{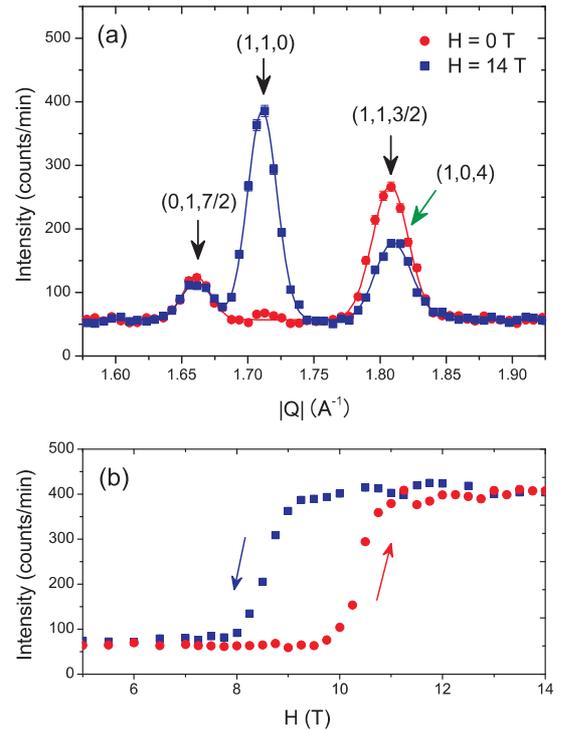}
\caption{(Color online) Elastic neutron scattering measurements on a polycrystalline sample of deuterated Ag jarosite were performed using the triple-axis spectrometer E1.  (a)~$|Q|$-scans at $T=3$ K show two magnetic Bragg peaks in zero field (red circles) at (1,~1,~3/2) and (0,~1,~7/2), and two magnetic peaks at $H=14$ T (blue squares) at (1,~1,~0) and (0,~1,~7/2).  The structural (1,~0,~4)  nuclear peak is denoted by the green arrow. (b)~The field dependence of the (1,~1,~0) peak intensity shows the hysteretic phase transition to the net $K_s\neq0$ state at high field.} \label{fig11}
\end{figure}

To further study the magnetic transition at $H_c$, elastic neutron scattering measurements on a polycrystalline sample in a high magnetic field were performed using the triple-axis spectrometer E1, the diffractometer E4 at the Berlin Neutron Scattering Center (BENSC), Helmholtz Centre Berlin for Materials and Energy (formerly Hahn-Meiter-Institut Berlin), with the incident neutron wavelength of 2.43 \AA~($\hbar\omega=13.9$ meV), and the triple-axis spectrometer BT7 at NCNR with the incident neutron energy of 14.7 meV. The horizontal collimations of $40^\prime-80^\prime-$S$-40^\prime-40^\prime$ and $40^\prime-$open$-$open were employed at E1 and E4, respectively, and the horizontal collimations of open$-50^\prime-$S$-60^\prime-120^\prime$ were employed at BT7. PG crystals were utilized to monochromate the incident neutron beam at E1, E4, and BT7, and to analyze the scattered neutron beam at E1 and BT7.  PG filters were placed in the incident beam at E1 and E4, and two PG filters were placed before and after the sample at BT7 to reduce higher-order contamination. At BENSC, deuterated polycrystalline Ag jarosite of mass 5.5 g was mounted inside the VM-1 magnet with the highest field of 14.5 T, and at NCNR 5.4~g of the polycrystalline sample was mounted inside the 11~T magnet. The sample was then cooled using a $^4$He cryostat. 

At zero field, the canted moments on successive kagome planes are equal and opposite. It takes six layers to form a magnetic unit cell instead of three layers for a structural unit cell (Figure~\ref{fig2}). This causes the doubling of a magnetic unit cell along the $c$-axis, resulting in magnetic Bragg peaks at a half integer value along the $00l$ direction. For $H>H_c$, the spins on alternate layers rotate 180$^\circ$ changing the spin arrangement from all-in to all-out and vice versa. Figure~\ref{fig2} shows such spin re-orientations on the second, fourth and sixth kagome planes. In this state, the magnetic unit cell contains only three kagome planes, and is the same size as the structural unit cell, resulting in the magnetic Bragg peaks at a whole integer value along the $00l$ direction. On the other hand, if the transition to the ferromagnetic state is caused by simply flipping the canted components on the alternate layers, the positions of the magnetic Bragg peaks will remain at the half integer value. 

\begin{figure}[t]
\centering
\includegraphics[width=2.8in]{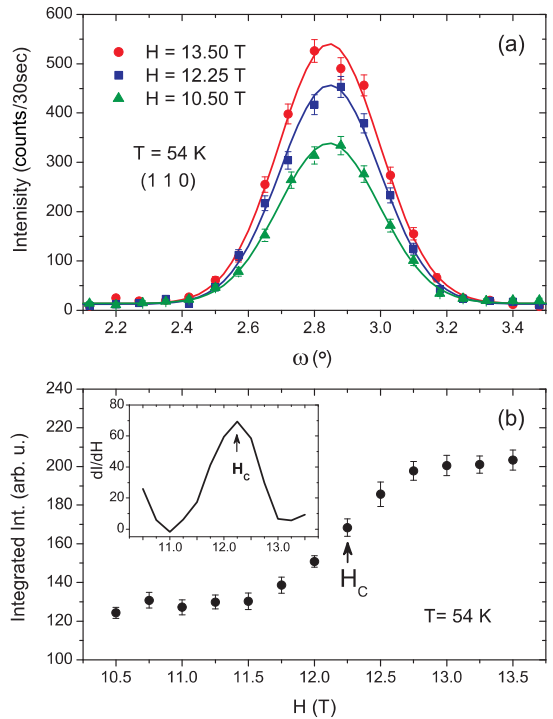}
\caption{(Color online) (a)~$\omega$-scans around (1, 1, 0) at three different magnetic fields at $T=54$ K measured at E1 on single-crystal K jarosite.  The magnetic field was applied along the $c$-axis, which is normal to the scattering plane.  The result shows the change in the integrated intensity of the (1, 1, 0) reflection as the field exceeds the critical field $H_c$.  (b)~The integrated intensity as a function of field measured at $T=54$ K shows a sudden change in intensity at the critical field.  The inset shows the derivative of the integrated intensity as a function of field.} \label{fig12}
\end{figure}

\begin{figure}[t]
\centering
\includegraphics[width=2.8in]{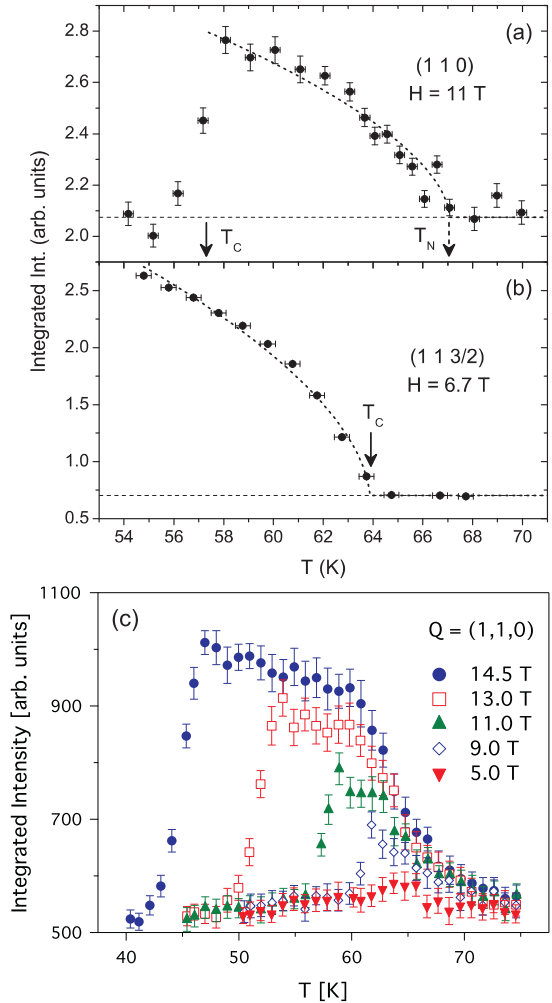}
\caption{(Color online) Neutron scattering intensities as a function of temperature were measured on a single-crystal sample of K jarosite (a)~at $Q=$(1, 1, 0) and $H=11$~T and (b)~ at $Q=$(1, 1, 3/2) and $H=6.7$~T.  A dotted arrow denotes a crossover phase line between the 3D ordered state with $K_s\neq0$ and the 2D short-range correlated, disordered state. Dotted lines are guides to the eye. (c) shows the field dependence of the (1,~1,~0) magnetic Bragg intensity.} \label{fig13}
\end{figure}

In Figure~\ref{fig11}, $|Q|$-scans at $T=3$~K measured on a deuterated polycrystalline sample of Ag jarosite at E1 reveal the shift of the magnetic Bragg peaks expected from the 180$^\circ$ spin rotation. To overcome the interlayer coupling and induce the 180$^\circ$ spin rotation, an applied field greater than the critical field is required along the crystallographic $c$-axis. At zero field, we observe two magnetic Bragg peaks at (0, 1, 7/2) and (1, 1, 3/2) in this $|Q|$ range. At $H=14$~T ($H>H_c$), the peak at (1, 1, 3/2) disappears, while a new magnetic peak at (1, 1, 0) emerges. The intensity at (0, 1, 7/2) does not change because the magnitude of the magnetic field component along the $c$-axis is less than the critical field. Since the applied field is perpendicular to the scattering plane, at (1, 1, 0) the $c$-axis is along the vertical direction, hence along the direction of the applied field. On the other hand, at (1, 1, 3/2) and (0, 1, 7/2) the angles between the field direction and the $c$-axis are 18$^\circ$ and 53$^\circ$, respectively.  Therefore, the field component along the $c$-axis is reduced by about 5\% at (1,~1,~3/2) and about 40\% at (0,~1,~7/2). With the applied field of 14~T, the field strengths along the $c$-axis at (1,~1,~3/2) and (0,~1,~7/2) are 13.3~T and 8.4~T, respectively, while the critical field at $T = 3$~K is 10.5~T. Thus, we observe a decrease in intensity at (1,~1,~3/2), but no change in intensity at (0,~1,~7/2) (Figure~\ref{fig11}(a)). The critical field was measured using the data taken at (1,~1,~0) where the applied field is perfectly along the $c$-axis, and the change in the intensity is largest.  The field dependence of the intensity at (1,~1,~0) (Figure~\ref{fig11}(b)) illustrates the hysteresis consistent with the magnetization data.  Furthermore, the critical fields obtained from the neutron scattering measurements are in good agreement with the magnetization results as shown in Figure~\ref{fig7}(b). 

To investigate this magnetic transition on a single-crystal sample, the elastic neutron scattering measurements on single-crystal K jarosite were performed using the triple-axis spectrometers E1 at BENSC and BT7 at NCNR with the same instrumental set-up as described above. Additional neutron diffraction was performed at the diffractometer E4 at BENSC with the incident neutron wavelength of 2.45 \AA~using a position sensitive detector. At BENSC, the single-crystal sample (mass $\approx$ 20~mg) was mounted inside the VM-1 magnet, and at NCNR the same sample was mounted inside the 11~T magnet.  The sample was cooled using a $^4$He cryostat. It was aligned in the $(hk0)$ plane with the magnetic field applied perpendicular to the scattering plane.  The intensity of the (1, 1, 0) reflection was measured as a function of field and temperature.  At NCNR, we measured the intensities of both (1, 1, 0) and (1, 1, 3/2) reflections to construct a complete phase diagram.  To measure the (1, 1, 3/2) reflection, the sample was re-aligned on the (2, 2, 3) structural Bragg peak in air, and the intensity of the (1, 1, 3/2) reflection was measured inside the 11~T magnet at low temperature.  As previously discussed, the magnetic field is along the $c$-axis for (1,1,0) and about 18$^\circ$ with respect to the $c$-axis for (1, 1, 3/2).  The values of magnetic fields in Figures~\ref{fig12} and~\ref{fig13} indicate the component of the field along the $c$-axis.

\begin{figure}[t]
\centering
\includegraphics[width=3.2in]{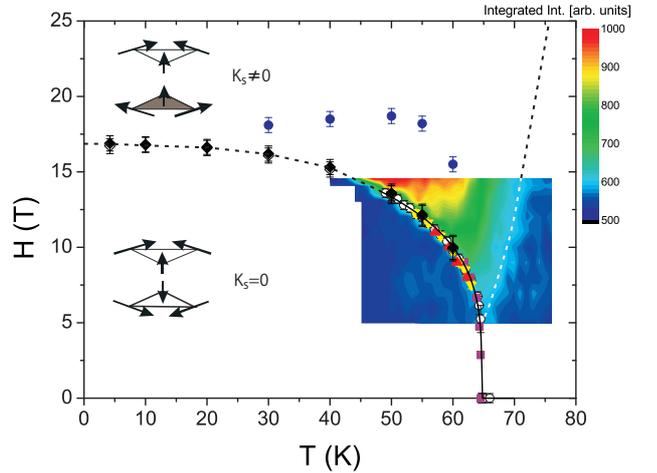}
\caption{(Color online) K jarosite phase diagram as a function of field and temperature.  Open circles show the results of the magnetization measurements.  Yellow down triangles show the results of the neutron scattering measurements at BENSC.  Red up triangles and purple squares show the results of the neutron scattering measurements at NCNR.  The critical fields obtained from the pulsed magnet at ISSP are denoted by closed and open diamonds for increasing and decreasing field sweeps, respectively.  Blue circles denote fields at which a kink in magnetization was observed as shown by black arrows in Figure~\ref{fig8}.  A contour plot shows the intensity of the (1,~1,~0) magnetic Bragg peak measured at E4. A solid line is a power law fit with exponent $\beta=0.25$.  Dotted lines, which serve as guides to the eye, denote the phase lines.} \label{fig14}
\end{figure}

Figure~\ref{fig12}(a) shows $\omega$-scans measured on a single-crystal sample at E1 around the (1,~1,~0) reflection at three different fields at $T =$ 54 K.  The data are fit to a Gaussian. Figure~\ref{fig12}(b) shows integrated intensity, which is extracted from the fit in Figure~\ref{fig12}(a), as a function of field at $T=$ 54 K.  The inset shows the derivative of the intensity, where the critical field is defined at the maximum.  The same measurements were performed at several temperatures to obtain $H_c(T)$. Yellow down triangles in Figure~\ref{fig14} show the critical field as a function of temperature, which is consistent with the magnetization measurements shown by open circles and diamonds.  Figures~\ref{fig13}(a) and~\ref{fig13}(b)  show neutron scattering intensities as a function of temperature and field measured at NCNR.  The magnetic scattering intensity at (1,~1,~0) (Figure~\ref{fig13}(a)) is non-zero in the ferromagnetic state, and goes to zero above $T_N$, where spins are disordered, and below $T_c$, where spins are in the antiferromagnetic state.  Therefore, by measuring the intensity at (1,~1,~0), one can extract both $T_N$ and $T_c$.  The intensity at (1,~1,~3/2) (Figure~\ref{fig13}(b)), on the other hand, is non-zero in the antiferromagnetic state, and goes to zero above $T_N$ or $T_c$. The arrows in Figures~\ref{fig13}(a) and~\ref{fig13}(b) indicate $T_c$ and $T_N$ from the (1,~1,~0) data and the (1,~1,~3/2) data.  In Figure~\ref{fig14}, $T_c(H)$ obtained from the (1,~1,~0) data is denoted by red up triangles, while that obtained from the (1,~1,~3/2) data is denoted by purple squares.  In addition, the field dependence of the (1,~1,~0) scattering intensity measured at E4 (Figure~\ref{fig13}(c) and a contour plot in Figure~\ref{fig14}) shows sharp drops on the low temperature side consistent with the spin re-orientation transition observed in the magnetization measurements.  On the high temperature side, a crossover between the 3D long-range ordered and 2D short-range correlated phases was observed.

\begin{figure}[t]
\centering
\includegraphics[width=2.85in]{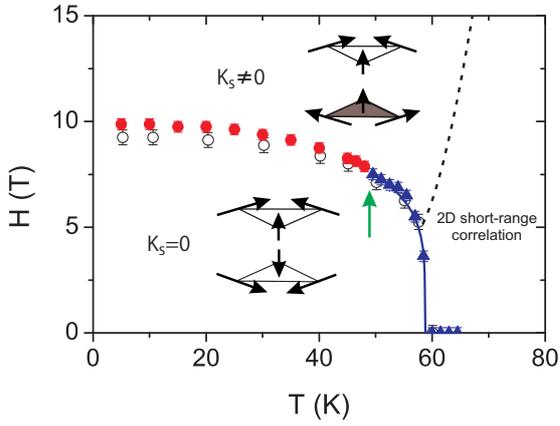}
\vspace{0mm} \caption{(Color online) Ag jarosite phase diagram as a function of field and temperature.  Red circles, denoting the first-order phase transition, show the average values of the critical fields from the magnetization measurements. Blue triangles, denoting the second-order phase transition, show the results of the magnetization measurements. Open circles show the NCNR neutron scattering results. A solid line is a power law fit with exponent $\beta=0.20$.  A dotted line shows an expected phase boundary between the net $K_s\neq0$ state and disordered state. A green arrow indicates the possibility of a multi-critical point.}\label{fig15}
\end{figure}

\section{Phase diagrams}

For Ag jarosite, the results of magnetization and neutron scattering measurements are summarized in the $H$-$T$ phase diagram shown in Figure~\ref{fig15}.  In the $K_s=0$ phase, the canted moments on adjacent kagome planes point in the opposite directions due to the weak interlayer coupling, resulting in a zero net magnetic moment.  However, at the critical field $H_c$, the Zeeman energy overcomes the interlayer coupling, and the spins on the alternate planes rotate 180$^\circ$, resulting in the reversal of the canted moments. In the magnetization measurements, the transition results in the abrupt increase of the magnetization at the critical field.  In the neutron scattering measurements, this transition shifts the magnetic Bragg peaks.

At low temperature, both magnetization and neutron scattering intensity show the hysteretic behavior as a function of field.  The hysteresis, which is indicative of the first-order transition, exists for $T\lesssim50$ K.  The magnetic transition line changes to second order for $T\gtrsim50$ K, raising a possibility of the multi-critical point at $T\approx50$~K. The hysteresis was also observed in K jarosite with a much smaller coercive field, which becomes too small to detect above 50 K.  However, we note that we cannot rule out the possibility that this first-order transition phase line could extend up to $T_N$.

The phase diagram of K jarosite is illustrated in Figure~\ref{fig14}.  The neutron scattering measurements show the field dependence of the 3D magnetic ordering temperature $T_N$.  At zero field, the 3D magnetic LRO is a result of the combined effect of the DM interaction and interlayer coupling.  The DM interaction induces the 2D long length-scale correlations in the planes, and the interlayer coupling propagates spin correlations in the third direction.  In the presence of a magnetic field, since the field favors the net $K_s\neq0$ state while the interlayer coupling favors the net $K_s=0$ state, they will compete with each other.  In the low field region, this competition leads to a decrease of $T_N$.  In the high field region, on the other hand, the Zeeman energy overwhelms the interlayer coupling, and takes over its role, resulting in an increase of $T_N$ as the field increases.  In Figure~\ref{fig14}, the contour plot of the (1,~1,~0) magnetic Bragg intensity measured at E4 shows a sharp transition at $(T_c, H_c)$, which corresponds to the spin re-orientation transition, and an indistinct phase line between the 3D ordered state with $K_s\neq0$ and the 2D short-range correlated, disordered state.  At $H\approx5$~T, the two phase lines meet, which is a special point where the effects of the interlayer coupling and magnetic field balance against each other, and the 3D magnetic LRO is solely induced by the 2D correlations in the planes. The lack of a well-defined peak in the specific heat at a finite field suggests that the phase boundary between the 3D $K\neq0$ ordered state and the 2D correlated, disordered state is a crossover rather than a true phase transition. The magnetic field applied along the $c$-axis breaks the scalar chirality symmetry, prohibiting further spontaneous breaking of the spin symmetry. This scalar chirality symmetry corresponds to the two oppositely canting degenerate states.  The spontaneous breaking of this $Z_2$ symmetry, which corresponds to the phase transition from the 2D short-range correlation state and the 3D long-rang magnetic ordered state, can only occur at zero field.  

\section{Conclusions}

We have investigated the spin canting due to the DM interaction in jarosite using specific heat, magnetization, and neutron scattering measurements.  From the observed critical field and canted moment obtained from our magnetization measurements, we were able to calculate the canting angle $\eta$, the DM parameters $|D_y|$ and $D_z$, and the interlayer coupling constant $J_\bot$. The in-plane component of the DM interaction forces the spins on a triangle to cant out of the kagome planes to form the umbrella structure of ferromagnetically aligned moments within the layers.  However, at zero field, the weak ferromagnetic interlayer coupling prompts the spins on neighboring planes to align in such an arrangement that the canted moments are equal and opposite.  At the critical field, the Zeeman energy overcomes the interlayer coupling causing the spins on the alternate layers to rotate $180^\circ$.  This $180^\circ$ spin rotation reverses the canted moments on the alternate layers causing the sudden increase in the magnetization. This field-induced spin re-orientation transition corresponds to the non-trivial change in the spin-texture. In particular, the transition yields a net, non-zero value for the scalar chirality. Our elastic neutron scattering measurements in a high magnetic field provide the first direct evidence of the 180$^\circ$ spin rotation at the transition. 

\acknowledgements 
The authors would like to thank P. Ghaemi and T. Senthil for many fruitful discussions. This work was supported by the NSF under Grant No. DMR 0239377, and in part by the MRSEC program under Grant No. DMR 02-13282.

\bibliography{DM_Ag_Jarosite}

\end{document}